\documentclass[12pt]{article}
\usepackage{graphicx}
\usepackage{epsfig}

\newcommand{\jpsi}{J/\Psi}
\newcommand{\be}{\begin{equation}}
\newcommand{\ee}{\end{equation}}

\newcommand{\bea}{\begin{eqnarray}}
\newcommand{\eea}{\end{eqnarray}}
\newcommand{\beq}{\begin{equation}}
\newcommand{\eeq}{\end{equation}}
\newcommand{\nn}{\nonumber}

\def\fun#1#2{\lower3.6pt\vbox{\baselineskip0pt\lineskip.9pt
\ialign{$\mathsurround=0pt#1\hfil##\hfil$\crcr#2\crcr\sim\crcr}}}

\begin{document}

\title{New narrow LHCb pentaquarks as lowest antiquark-diquark-diquark systems}

\author{A.N. Semenova, V.V. Anisovich, A.V. Sarantsev}
\maketitle

\begin{center}
{\it Petersburg Nuclear Physics Institute of National Research Centre ''Kurchatov Institute'', Gatchina, 188300, Russia}
\end{center}

\begin{abstract}
The antiquark-diquark-diquark model describes pentaquark
states both in terms of quarks and hadrons. We discuss pentaquark
states with hidden charm $P(\bar c cuud)$ discovered in the $J/\Psi
p$ spectrum by the LHCb collaboration. We consider three pentaquark
states as members of the lowest ($S$-wave) multiplet and discuss the
mass splitting scheme. The latest LHCb data for pentaquarks with
hidden charm provide an opportunity to make an assumption about the
diquark content of the pentaquark states. We give a classification
for the LHCb pentaquarks and define recombination channels for these
states.
\end{abstract}



\section{Introduction}
The observation of the pentaquark state  $P_c(4450)^+$ in the
reaction $\Lambda_b\to J/\Psi K^- p$ reported by the LHCb
collaboration was one of the most essential issues in the hadron
spectroscopy \cite{Lhcb}. This state was associated with a
relatively narrow peak in the $J/\Psi p$ channel, however the
partial wave analysis of the measured data could not identify
uniquely it's quantum numbers. Recently the LHCb collaboration
collected the new data and performed the particle identification
analysis. The total statistic was increased by nine times and the
detection efficiency of the $\Lambda_b$ state in the new data set
was notably improved. The new analysis showed that the earlier
observed signal has a two peak structure and is identified at
present as the contribution from the two pentaquark states
$P_c(4440)^+$ and $P_c(4457)^+$ \cite{LHCbn}. Moreover the new data
allowed the collaboration to observe a new state $P_c(4312)$ which
escaped the identification in the low statistic data. However, in
spite of the very high statistic of the new data the LHCb
collaboration was not able to define the quantum numbers of the
observed states.

The nature of the observed states was intensively discussed since
the first LHCb report was published. The treatment of pentaquark as
a five quark bound state produces a huge numbers of the resonances
in the light and heavy quark sector. Therefore it is clear that a
pentaquark state should have a configuration with much less degrees
of freedom. One of the popular suggestion is that the pentaquark is
the baryon-meson molecule. This idea was discussed in the original
LHCb \cite{LHCbn} and was based on the observation that the found
states are situated near the thresholds of $\Sigma_c$ hyperons and
$D(D^*)$ mesons. This idea was investigated in the set of papers,
see \cite{Peng:2019wys}-\cite{Lin:2019qiv} and references therein.
In most of these papers the lowest pentaquark was suggested to be
the $\Sigma_c\bar D$ loosely bound states with binding energy 5 MeV
while the two higher mass states were suggested to be $\Sigma_c\bar
D^*$ molecules with binding energy 2 and 20 MeV. However in many
cases the resonances situated near thresholds have zero (or a very
small) couplings to the corresponding channels due to specific
resonance quantum numbers. Therefore until the quantum numbers of
the observed states are identified it is difficult to believe that
the correct interpretation is found. Moreover the different models
predict the different resonance spectrum. For example, in the papers
\cite{Peng:2019wys},\cite{Xiao:2019aya} apart of the observed states
which were explained as the $\Sigma_c\bar D$ and $\Sigma_c\bar D^*$
bound systems three additional states were predicted: one state with
the mass around 4375 MeV and two states with the mass around 4520
MeV. Another model \cite{Chen:2019bip} suggested that the observed
states can be not only $\Sigma_c\bar D$ and $\Sigma_c\bar D^*$ bound
states but also $\Sigma_c^*\bar D$ and/or $\Sigma_c^*\bar D^*$
molecules depending on their quantum numbers. For our opinion the
treatment of the observed states as $\Sigma^{(*)}_c D^{(*)}$ bound
states has a problem. In the initial $\Lambda_b$ hyperon the
nonstrange quarks are in the isoscalar configuration. This explains
why in the $Kp$ channel the LHCb observed the dominant production of
the $\Lambda$ hyperons but not a production of $\Sigma$ hyperons
(isospin is not conserved for the weak interactions). However the
$\Sigma_c$ hyperons have the nonstrange quarks in the isovector
state which suggests a severe quark recombination. Therefore it is
difficult to expect the production of the $\Sigma^{(*)}_c D^{(*)}$
molecules from the $\Lambda_b$ decay. It is pity that all papers
which put forward the molecular picture do not discuss the
production mechanism of such states from the $\Lambda_b$ decay.

Another interpretation is based on the idea that the pentaquarks are
bound states of a heavy quarkonium and a light baryon (see
Ref.~\cite{Eides:2019tgv} and references therein). Such states are called
hadroquarkonium and their binding energy is order of the hundred
MeV.

The idea which naturally decreases the number of resonances is a
formation of diquark states with the $L=0$ orbital momentum between
quarks. In this case the pentaquark is a $\bar c\cdot (cu)\cdot
(ud)$ bound system (see
Refs.~\cite{wang,maia-polo,ala-2,ccstr,mpla17}). In  the present
paper we also discuss this idea which can naturally explain the
observed spectrum and predicts exactly three states observed by the
experiment. In this case the quantum numbers of the lowest
pentaquark state $P_c(4312)^+$ should have the $J^P=1/2^-$ quantum
numbers and the $P_c(4440)^+$ and $P_c(4457)^+$ states should have
spin-parity $1/2^-$ and $3/2^-$. According to the
antiquark-diquark-di\-quark model all these states are members of
the same $P=\bar c\cdot (cu)\cdot (ud)$ multiplet. In the present
paper we discuss the possible configuration of the observed states
and calculate their recombination transitions. We suppose that this
recombination defines the main part of the decay process of the
observed states.

\section{Color-spin-isospin structure of the pentaquark}

With $L=0$ orbital moment two kinds of diquark can be formed: the
scalar diquark $S^{II_z,JJ_z}$ and the axial-vector diquark
$A^{II_z,JJ_z}$. Such diquarks can be formed either between two
light quarks $q_i$ or between the heavy and light quarks:
\bea
\label{5}
&&
S_{cq}^{\frac 12 I_z,00}(I=1/2,\,J=0),\qquad
A_{cq}^{\frac 12 I_z,1J_z}(I=1/2,\,J=1),\nn
\\
&& S_{q_1q_2}^{00,00}(I=0,\,J=0),\qquad\,\,
A_{q_1q_2}^{1I_z,1J_z}(I=1,\,J=1),
\eea
where $I$ and $J$ refer to isospin and spin of the diquarks.

In terms of the antiquark-diquark-diquark system pentaquark can be presented as a three-body system:
\bea \label{2mn}
P=a\;(\bar c^{\alpha}\cdot   D_{cq_1}^{\beta}\cdot D_{q_2q_3}^{\gamma}\epsilon_{\alpha\beta\gamma})
+b\;(-\bar c^{\alpha}\cdot  D_{q_2q_3}^{\beta}\cdot
D_{cq_1}^{\gamma}\epsilon_{\alpha\beta\gamma}),
\eea
where $\alpha,\beta,\gamma$ are the color indices. We use '$\cdot$'
symbol to separate the compact configurations. The two terms
describe the decay of the system due to the decomposition of the
heavy diquark (first term) and the light diquark (second term). The
decay rate for such decomposition is described by the coefficients
$a, b$. The diquarks have the same color structure as antiquarks:
\bea \label{21mn}
D_{cq_1}^{\beta}& =& c_{\beta'}q_{1\beta''}\epsilon^{\beta\beta'\beta''},
D_{q_2q_3}^{\gamma} = q_{2\gamma'}q_{3\gamma''}\epsilon^{\gamma\gamma'\gamma''}.
\eea
The second term in Eq.~(\ref{2mn}) gives a zero result for the states under consideration. Taking into account coordinate wavefunction lead us to nonvanishing second term in Eq.~(\ref{2mn}), however such states are radial excited ones and they are out of our consideration. Substituting (\ref{21mn}) into (\ref{2mn}) and convoluting color indecies one can obtain the  recombination of a pentaquark into a meson-baryon channels:
\bea\label{22nm}
&&-\bar c^{\alpha}\cdot  D_{q_2q_3}^{\beta}\cdot D_{cq_1}^{\gamma}\epsilon_{\alpha\beta\gamma}=\frac{1}{\sqrt 2}({\bar c}^{\alpha}q_{2\alpha})(q_{3\beta}c_{\beta'}q_{1\beta''}
\epsilon^{\beta\beta'\beta''})
-\frac{1}{\sqrt 2}({\bar c}^{\alpha}q_{3\alpha})(q_{2\beta}c_{\beta'}q_{1\beta''}
\epsilon^{\beta\beta'\beta''})=0.
\eea
The coefficient $1/\sqrt 2$ is the normalizing one. The zero results
due to flavor-spin symmetry of the light diquark:
$D_{q_2q_3}=D_{q_3q_2}$. Let us consider as an example the
combination $P=\bar c \cdot A_{u^{\uparrow}d^{\uparrow}}\cdot
A_{c^{\uparrow}u^{\uparrow}} $:
\bea
&&
\bar c^{\uparrow}\cdot  A_{u^{\uparrow}d^{\uparrow}}\cdot A_{c^{\uparrow}u^{\uparrow}}
= \bar c^{\uparrow}\cdot \frac{1}{\sqrt 2}(u^{\uparrow}d^{\uparrow}+
d^{\uparrow}u^{\uparrow})\cdot
c^{\uparrow}u^{\uparrow}=
\\
&&
= \frac{1}{\sqrt 2} \Big(
\bar c^{\uparrow}\cdot  u^{\uparrow}d^{\uparrow}\cdot  c^{\uparrow}u^{\uparrow}
+
\bar c^{\uparrow}\cdot d^{\uparrow}u^{\uparrow} \cdot  c^{\uparrow}u^{\uparrow}
\Big)=
\nn
\\
&&
= \frac{1}{\sqrt 2} \Big(
-\bar c^{\uparrow}u^{\uparrow}\cdot  d^{\uparrow} c^{\uparrow}u^{\uparrow}
+\bar c^{\uparrow}d^{\uparrow}\cdot u^{\uparrow}  c^{\uparrow}u^{\uparrow}-\bar c^{\uparrow}d^{\uparrow}\cdot  u^{\uparrow} c^{\uparrow}u^{\uparrow}
+\bar c^{\uparrow}u^{\uparrow}\cdot  d^{\uparrow}c^{\uparrow} u^{\uparrow}
\Big)=0.
\nn
\eea
So we need to consider only the first term in Eq.~(\ref{2mn}).  And
the light diquark $D_{q_2q_3}$ will form the final baryon state.
This is a very important phenomenon. In the initial $\Lambda_b$
hyperon the two light quarks form the isoscalar system and if this
system is transferred to the pentaquark state it can not decay into
the channel with $\Sigma_c$ hyperon and meson due to quark
recombination.

In terms of scalar and axial diquarks (\ref{5}) the color-flavor wave function of pentaquark
reads:
\be
\label{6}
P_{\bar c\cdot cq\cdot q_1q_2}=\bar c^\alpha\cdot \epsilon_{\alpha\beta\gamma}\,
\begin{tabular}{|l|}
$S^\beta_{cq_1}$\\
$A^\beta_{cq_1}$
\end{tabular}
\cdot
\begin{tabular}{|l|}
$S^\gamma_{q_2q_3}$\\
$A^\gamma_{q_2q_3}$
\end{tabular}
,\ee where $q_i$ refer to light quarks $u$, $d$, indices $\alpha$,
$\beta$, $\gamma$ refer to color. So, the low-lying $S$-wave
pentaquark multiplet $P(I,J^P)$ for isospin $I=\frac{1}{2}$ reads:
\be\label{7a}
P_{\bar c\cdot (cq_1)\cdot(q_2q_3)}=
\begin{tabular}{l}
$P_{S_cS}(\frac12,\frac12^{-})$,\\
$P_{S_cA}(\frac12,\frac12^{-})$, $P_{S_cA}(\frac12,\frac32^{-})$,\\
$P_{A_cS}(\frac12,\frac12^{-})$, $P_{A_cS}(\frac12,\frac32^{-})$,\\
$P_{A_cA}(\frac12,\frac12^{-})$,\\
$P_{A_cA}(\frac12,\frac12^{-})$, $P_{A_cA}(\frac12,\frac32^{-})$,\\
$P_{A_cA}(\frac12,\frac32^{-})$, $P_{A_cA}(\frac12,\frac52^{-})$.\\
\end{tabular}
 \ee
As we can see the consideration of four diquarks leads to ten
pentaquark states.  The scalar diquark $S$ is a "good diquark" in
Wilczek and Jaffe's terminology and has a lower energy than the
axial one.

\section{Mass splitting scheme}\label{sec1}

All three observed pentaquarks with hidden charm are relatively
narrow states and have a specific mass patten. The mass splitting
and decay rate was discussed in the number of papers
\cite{wang,Petrov,espo,os,pol,ali,skw,wo,st}.

Let us consider the mass splitting  scheme for the observed
pentaquarks on the basis of the hypothesis about their quark-diquark
nature. As it was shown in Refs.~\cite{ZSah,ruh-gl,GL} the mass
splitting of hadrons can be well described in the framework of the
quark model by the short-ranged color-magnetic interactions of the
constituents. For mesons and baryons the mass formulae for the
ground states was obtained by Glashow \cite{GL}:
\bea \label{7}
&& M_M=\sum\limits_{j=1,2}m_{q(j)}+\tilde a
\frac{\vec{s}_1\vec{s}_2}{m_{q(1)}m_{q(2)}},
\\
&& M_B=\sum\limits_{j=1,2,3}m_{q(j)}+\tilde b
\sum\limits_{j>\ell}\frac{\vec{s}_j\vec{s}_\ell}{m_{q(j)}m_{q(\ell)}},
\nn
\eea
where  $\vec{s}_j$ and  $m_{q(j)}$ refer to spins and masses of the
constituents. Mass splitting parameters in Eq.~(\ref{7}) $\tilde a$
and $\tilde b$ are characterized by the size of the color-magnetic
interaction and the size of the formed hadrons, the source of the
short-range interaction was suggested in Ref.~\cite{GL}. Initially
equations (\ref{7}) were put forward for the 36-plet of mesons
($q\bar q$) and 56-plet of baryons ($qqq$) and works there very
well.

Modified formulae (\ref{7}) can be naturally applied to the
pentaquark systems. Here we use the idea that diquark-quark
interaction has a similar nature as antiquark-quark interaction.
Then the pentaquark can be considered as antibaryon state. For
multiquark states from Ref.~\cite{LHCbn} we have:
\bea \label{8}
&&M_{q_1q_2\cdot q_3c\cdot \bar c}=m_{D(q_1q_2)D(q_3c)\bar c}+4\Delta\Big(\vec{\mu}_{D(q_1q_2)}\vec{\mu}_{D(q_3c)}+
\vec{\mu}_{D(q_1q_2)}\vec{\mu}_{\bar c}+\vec{ \mu}_{\bar c}\vec{\mu}_{D(q_3c)}\Big),
\nn \\
&&m_{D(q_1q_2)D(q_3c)\bar c}=
m_{D(q_1q_2)}+m_{D(q_3c)}+m_{\bar c}
\eea
where $\vec \mu_D$ and $\vec \mu_{\bar c}$ are color-magnetic
moments of diquarks and $c$-quark, $\Delta$ is the parameter of spin
splitting. The magnetic moments are written as sums of quark
magnetic moments:
\bea
\label{10e}
&&
\vec{\mu}_{D(q_1q_2)}=\vec{s}_{q_1}\frac{m_{q}}{m_{q}}+
\vec{s}_{q_2}\frac{m_{q}}{m_{q}},\nn\\
&&
\vec{\mu}_{D(q_3c)} = \vec{s}_{q_3}\frac{ m_{q}}{m_{q}}
+\vec{s}_c \frac{m_{q}}{m_{c}}\simeq \vec{s}_{q_3} ,\nn\\
&& \vec{\mu}_c=\vec{s}_c\frac{m_q}{m_c}\simeq 0.
\eea
Here we take into account that $m_{c}>>m_{q}$ and we follow the idea
that the spin-spin interaction is suppressed for heavy quarks
\cite{ZSah}.

\subsection{Estimation of diquark and pentaquark masses}

Estimation of  diquark masses is the most problematic issue in the
study of diquarks (see for example Ref.~\cite{santopinto}). Basing
on Refs.~\cite{lightdiq}, \cite{QQQQ} we estimate the masses of
scalar $S$ ($J^P=0^+$) and axial $A$ ($J^P=1^+$) diquarks as follows
(in MeV units):
\be\label{es}
\begin{tabular}{ll}
$m_q=330,$ & $m_c=1450$,\\
$m_{S(q_1q_2)}=750$, & $m_{S(q_3c)}=2100$,\\
$m_{A(q_1q_2)}=850$, & $m_{A(q_3c)}=2200$.\\
\end{tabular}
\ee
Here we explore the most general patten with existence of the axial
diquark between heavy and the light quarks.

As it would be discussed below it is natural to consider for LHCb
pentaquarks two configurations: the first one consisting of two
scalar diquarks $P_c=\bar c S_cS$ and the second one consisting of
one scalar and one axial diquark $P_c=\bar c S_cA$, $P_c=\bar c
A_cS$. Then in the mass region below 4500 MeV we obtain the five
states with following masses:
\be
{\begin{tabular}{@{}c|cc@{}}
$ P_{DD_c\bar{c}}^{(L,J^{P})}$& mass  &MeV \\
\hline
$P_{SS_c\bar{c}}^{(0,\frac12^{-})}$ & $m_{SS_c\bar{c}} $ & $\simeq 4300$\\
\hline
$P_{AS_c\bar{c}}^{(1,\frac12^{-})}$ & $m_{AS_c\bar{c}}$ &  $\simeq 4400$ \\
$P_{AS_c\bar{c}}^{(1,\frac32^{-})}$& $m_{AS_c\bar{c}}$ & $\simeq 4400$ \\
\hline
$P_{SA_c\bar{c}}^{(1,\frac12^{-})}$&$m_{SA_c\bar{c}}$&$\simeq 4400$\\
$P_{SA_c\bar{c}}^{(1,\frac32^{-})}$&$m_{SA_c\bar{c}}$&$\simeq 4400$\\
\end{tabular}} \label{ta1}
\ee
Really in the Glashow formulae  the coordinate part of the wave
function is hidden in parameters $a$ and $b$. They may be different
for standard and exotic hadrons.  Therefore we should emphasize that
all our mass estimations are only qualitative ones.

\section{Production mechanism for LHCb pentaquarks}

The diquark picture suggests five states in the mass region below
4500 MeV. The pentaquarks which formed by the two axial diquarks
should have masses in the region 4600 MeV. However in a particular
reaction some of the states can be forbidden (or have a very small
probability) due to the production mechanism. The LHCb pentaquark
states were observed in the decay of the $\Lambda_b$ meson into
$J/\Psi K^-p$ system. Such decay is defined by the weak transition
of the $b$ quark into $c\bar c s$ system. The $\Lambda_b$ is formed
by the heavy $b$ quark and the light scalar diquark and it is
natural to suggest that this diquark forms the final pentaquark
states. In this case we expect that only three states can be
observed in the reaction $\Lambda_b\to J/\Psi K^-p$:
$P_{SS_c\bar{c}}^{(0,\frac12^{-})}$,
$P_{SA_c\bar{c}}^{(1,\frac12^{-})}$ and
$P_{SA_c\bar{c}}^{(1,\frac32^{-})}$. The pentaquark with light axial
diquark should be produced in the decay of the $\Sigma_b$ particle.
However this is a very rare event due to dominant decay of the
$\Sigma_b$ baryon into $\Lambda_b\pi$ system. Possibly such states
can be seen in the reaction of the proton-antiproton annihilation
which will be studied by the PANDA experiment.

\section{Recombination channels}
Let us present the recombination scheme for the discussed
pentaquarks. We use the notation $P^{II_z,JJ_z}$, where $I$ and $J$
refer to isospin and spin of the pentaquark correspondingly.

We start with the Eq.~(\ref{6}). Taking into account the color
structure of the pentaquark and the diquark
$D^{\alpha}_{q_1q_2}=q_1^{\beta}q_2^{\gamma}\epsilon_{\alpha\beta\gamma}$
one can obtain recombination of the pentaquark into the white
meson-baryon channels:
\bea\label{3m}
&&
P_{\bar c\cdot cq\cdot q_1q_2}=\bar c^{\alpha}\cdot   D_{cq_1}^{\beta}\cdot D_{q_2q_3}^{\gamma}
\epsilon_{\alpha\beta\gamma}=\\
&=&-\frac{1}{\sqrt 2}({\bar c}^{\alpha}c_{\alpha})
(q_{1\gamma}q_{2\gamma'}q_{3\gamma''}\epsilon^{\gamma\gamma'\gamma''})+\frac{1}{\sqrt 2}
({\bar c}^{\alpha}
q_{1\alpha})(c_{\gamma}q_{2\gamma'}q_{3\gamma''}\epsilon^{\gamma\gamma'\gamma''})=
\nn \\
&=&
-\frac{1}{\sqrt 2}({\bar c}c)(q_{1}q_{2}q_{3})+\frac{1}{\sqrt 2}({\bar c}q_{1})(cq_{2}q_{3})
\nn.
\eea
Here the last line presents the result in a short form, the brackets
separate the white meson and baryon states. To obtain spin and
isospin quantum numbers of mesons and baryons at the end of
Eq.~(\ref{3m}) one need to substitute the exact expressions for the
diquarks $D$ in the beginning of Eq.~(\ref{3m}).

Now let us consider the pentaquark states with the different
diquark-diquark content listed in Eq.~(\ref{ta1}).

In the simplest case $P=\bar cS_{cu}S_{ud}$ we will provide some
details of the calculations. For two scalar diquarks: the heavy one
$S_{cu}^{\frac12\frac12,00}=[c^{\uparrow}u^{\downarrow}-c^{\downarrow}u^{\uparrow}]\frac{1}{\sqrt
2}$ and the light one
$S_{ud}^{00,00}=[u^{\uparrow}d^{\downarrow}-u^{\downarrow}d^{\uparrow}-d^{\uparrow}u^{\downarrow}+d^{\downarrow}u^{\uparrow}]\frac{1}{2}$
(the arrows denote the projection of spin) we obtain the following
recombination scheme:
\bea \label{n1}
&&P^{\frac12\frac12,\frac12\frac12}_{\bar c\uparrow S_{cu}S_{ud}}=
\bar c^{\uparrow\alpha}[c^{\uparrow}u^{\downarrow}-c^{\downarrow}u^{\uparrow}]^{\beta}\frac{1}{\sqrt 2}S_{ud}^{\gamma}\epsilon_{\alpha\beta\gamma}= \\
&&=\frac{1}{ 2}\left\{-\bar c^{\uparrow}c^{\uparrow}\cdot u^{\downarrow}S_{ud}+\bar c^{\uparrow}u^{\downarrow}\cdot c^{\uparrow}S_{ud}+\bar c^{\uparrow}c^{\downarrow}\cdot u^{\uparrow}S_{ud}-\bar c^{\uparrow}u^{\uparrow}\cdot c^{\downarrow}S_{ud}\right\}.\nn
\eea
Rewriting $(\bar qq)$ and $(qD)$ combinations in terms of mesons and baryons (see.~ Appendix \ref{appA}) we obtain:
\bea\label{n4}
&&P^{\frac12\frac12,\frac12\frac12}_{\bar c\uparrow S_{cu}S_{ud}}=+\frac{1}{4}(J/\Psi^{(0)}-\eta_c)(p^{\uparrow}+p^{'\uparrow})-\frac{1}{2}\bar D^{*0\Uparrow}\Lambda_c^{+\downarrow}-\nn\\
&&-\frac{1}{2\sqrt 2}J/\Psi^{\Uparrow}(p^{\downarrow}+p^{'\downarrow})+\frac{1}{2\sqrt 2}(\bar D^{*0(0)}-\bar D^{0})\Lambda_c^{+\uparrow}.
\eea
Here $p'$ is some radial excitation of proton. In the mass region of
the observed pentaquarks the open channels are $J/\Psi p$, $\bar
D^0\Lambda_c$ and $\eta_c p$.  Therefore the lowest mass state
$P_c(4312)^+$ should be also seen in the $\bar D^0\Lambda_c$ and
$\eta_c p$ channels. However the production rate in these channels
will be suppressed compare to the $J/\Psi p$ channel by the factors
0.6 and 0.4 correspondingly.

For the case of the axial diquark formed by the heavy and light
quark and the scalar light diquark we have the following result:
\bea\label{ac1}
&&P_{\bar cA(cu)S(ud)}^{\frac12\frac12,\frac12\frac12}=\big[-\frac{1}{\sqrt 6}(J/\Psi^{(0)}+\eta_c)(p^{\uparrow}+p^{'\uparrow})+\frac{1}{\sqrt 3}(\bar D^{*0(0)}+\bar D^0)\Lambda_c^{+\uparrow}-\frac{1}{\sqrt 6}\bar D^{*0\Uparrow}\Lambda_c^{+\downarrow}+\nn\\
&&+\frac{1}{\sqrt{12}}J/\Psi^{\Uparrow}(p^{\downarrow}+p^{'\downarrow})-\frac{1}{\sqrt{12}}(\bar D^{*0(0)}-\bar D^0)\Lambda_c^{+\uparrow}+\frac{1}{2\sqrt
6}(J/\Psi^{(0)}-\eta_c)(p^{\uparrow}+p^{'\uparrow})\big]\frac{1}{\sqrt2}.
\eea

\bea\label{ac3}
P_{\bar
cA(cu)S(ud)}^{\frac12\frac12,\frac32\frac32}=\frac{1}{\sqrt 2}\bar
D^{*0\Uparrow}\Lambda_c^{+00,\frac12\frac12}-\frac{1}{2}J/\Psi^{\Uparrow}(p^{\uparrow}+p^{'\uparrow}).
\eea
The state with quantum numbers $J^P=3/2^-$ does not decay into the
$\bar D^0\Lambda_c$ and $\eta_c p$ final states in the $S$-wave. It
can decay into these channels in the $D$-wave, however such production
should be heavily suppressed by the small phase volume.
\be
{\begin{tabular}{|c|ccc|}
~            & $\jpsi p$ &$\eta_c p$ &$\bar D^0\Lambda_c$\\
$P_c(4312)$  & 3/8$\times$1      &  1/8$\times$1.20 &  1/8$\times$0.88 \\
$P_c(4440)$  & 1/8$\times$1      &  3/8$\times$1.14 &  3/8$\times$0.97 \\
$P_c(4457)$  & 1/2$\times$1      &  0 &  0 \\
\end{tabular}} \label{br}
\ee

The $S$-wave phase volumes are very similar for the $J/\Psi p$, $\bar
D_0\Lambda_c$ and $\eta_c p$ channels. The product of the squared
decay coefficients from Eqs.~(\ref{n4}-\ref{ac3}) and phase volumes
(normalized to the $\jpsi p$ phase volume) are given in
Eq.~(\ref{br}). Therefore the production rate of the high mass
$1/2^-$ state will be by the factor 2.9 larger for the $\bar
D^0\Lambda_c$ decay channel and by the factor 3.4 larger for the
$\eta_c p$ channel compare to the $J/\Psi p$ channel. Moreover in
the $\bar D^0\Lambda_c$ and $\eta_c p$ final states only two signals
from the $J^P=1/2^-$ resonances will be observed. If we assume that
the two $1/2^-$ states are produced in the $\jpsi p$ channel with a
similar strength then in the $\bar D_0\Lambda_c$ and $\eta_c p$
channels the second state will be produced be by almost 10 times
stronger.

 For the case of heavy scalar diquark and light axial one $P=\bar c
S_{cu}\cdot A_{ud}$ we have:
\bea\label{au1}
&&P_{\bar cS(cu)A(ud)} ^{\frac12\frac12,\frac12\frac12}=\big[-\frac{\sqrt 2}{3}J/\Psi^{(0)}N^{\frac12\frac12,\frac32\frac12}-\frac{\sqrt 2}{12}J/\Psi^{(0)}p^{\uparrow}+\frac{\sqrt 2}{12}J/\Psi^{(0)}p^{'\uparrow}+\frac{\sqrt 2}{4}\eta_c p^{'\uparrow}+\nn\\
&&+\frac{\sqrt 3}{3}J/\Psi^{\Downarrow}N^{\frac12\frac12,\frac32\frac32}+\frac13 J/\Psi^{\Uparrow}N^{\frac12\frac12,\frac32-\frac12}+\frac16 J/\Psi^{\Uparrow}p^{\downarrow}-\frac16 J/\Psi^{\Uparrow}p^{'\downarrow}-\frac{\sqrt 2}{4}\eta_c p^{\uparrow}-\nn\\
&&-\frac{\sqrt 6}{18}\bar D^{*0\Uparrow}\Sigma_c^{+10,\frac12-\frac12}-\frac13\bar D^{*0\Downarrow}\Sigma_c^{+10,\frac32\frac32}+\frac{\sqrt 3}{6}\bar D^0\Sigma_c^{+10,\frac12\frac12}-\frac{\sqrt 3}{9}\bar D^{*0\Uparrow}\Sigma_c^{+10,\frac32-\frac12}+\\
&&+\frac{\sqrt 3}{18}\bar D^{*0(0)}\Sigma_c^{+10,\frac12\frac12}
+\frac{\sqrt 6}{9}\bar D^{*0(0)}\Sigma_c^{+10,\frac32\frac12}+\frac{\sqrt 3}{9}D^{*-\Uparrow}\Sigma_c^{++11,\frac12-\frac12}-\frac{\sqrt 6}{6}D^-\Sigma_c^{++11,\frac12\frac12}-\nn\\
&&-\frac{\sqrt 6}{18}D^{*-(0)}\Sigma_c^{++11,\frac12\frac12}+\frac{\sqrt 6}{9}D^{*-\Uparrow}\Sigma_c^{++11,\frac32-\frac12}-\frac{2\sqrt 3}{9}D^{*-(0)}\Sigma_c^{++11,\frac32\frac12}+\frac{\sqrt 2}{3}D^{*-\Downarrow}\Sigma_c^{++11,\frac32\frac32}\big]\frac{1}{\sqrt2}.\nn
\eea
Here the baryon $N^{\frac12\frac12,\frac32 J_Z}$ is the well known
$N(1520)3/2^-$.

For the pentaquark states with spin and its projection
$JJ_z=\frac{3}{2}\frac{3}{2}$ we obtain:
\bea\label{au3}
&&P_{\bar cS(cu)A(ud)}^{\frac12\frac12,\frac32\frac32}=\frac{1}{6\sqrt2}[3(J/\Psi^{(0)}-\eta_c)N^{\frac12\frac12,\frac32\frac32}-\sqrt{6}J/\Psi^{\Uparrow}(N^{\frac12\frac12,\frac32\frac12}+p^{\uparrow}-p^{'\uparrow})+\nn\\
&&+\bar D^{*0\Uparrow}(2\Sigma_c^{+10,\frac12\frac12}+\sqrt{2}\Sigma_c^{+10,\frac32\frac12})-\sqrt{3}(\bar D^{*(0)}-\bar D^0)\Sigma_c^{+10,\frac32\frac32}-\nn\\
&&-2D^{*-\Uparrow}(\sqrt{2}\Sigma_c^{++11,\frac12\frac12}-\Sigma_c^{++11,\frac32\frac12})+\sqrt{6}(D^{*-(0)}-D^-)\Sigma_c^{++11,\frac32\frac32}],
\eea
As we can see such states can be observed from their decay into
$J/\Psi p$ and $\eta_c p$ channels. However the states with the
axial light diquark should be produced from the decay of the
$\Sigma_b$ state, where the weak decay is strongly suppressed by the
strong decay into the $\Lambda_b\pi$ channel.

\section{Conclusion}

We  discuss the description of three LHCb pentaquarks in terms of
diquark-diquark-antiquark model. Basing on the idea that the light
scalar diquark which forms initial $\Lambda_b$ state directly
participates in the formation of the final pentaquark state we
reproduce the mass spectrum observed by the LHCb collaboration. The
presence of the initial isoscalar configuration naturally explains
why the $\Lambda$ hyperons are produced dominantly in the $Kp$
channel. If this configuration directly participates in formation of
the pentaquark states, these pentaquarks can not decay into the
$\Sigma_c^{(*)}\bar D^{(*)}$ channel due to quark recombination
mechanism. So the correspondence to the thresholds is an accident
in the leading order.

In our model the lowest LHCb state $P_c^+(4312)$ is formed by two
scalar diquarks and has the structure ${P=\bar c S_{cu}\cdot
S_{ud}}$ with quantum numbers $I,J^P=1/2,1/2^-$. The two states
$P_c^+(4440)$ and $P_c^+(4457)$ are formed by the axial heavy
diquark and the light scalar diquark and have quantum numbers
$I,J^P=1/2,1/2^-$ and $I,J^P=1/2,3/2^-$. The states with quantum
numbers $I,J^P=1/2,1/2^-$ can be also observed in the $\bar
D^0\Lambda_c$ and $\eta_c p$ decay channels. Here the production
rate of the lowest mass state into these channels will be suppressed
by the factors 0.3 and 0.4 (compare to the decay rate into $J/\Psi
p$) while the production rate of the high mass state will be
increased by the factors 2.9 and 3.4 correspondingly.

The pentaquark states with the axial light diquark can be produced
form the weak decay of the $\Sigma_b$ state. However such reaction
has a very small decay rate due to possibility of the $\Sigma_b$
hyperon to decay strongly into the $\Lambda_b\pi$ state. One of the
possible sources to observe such states is the proton-antiproton
annihilation reaction or the $\gamma p$ collision data.

\section*{Acknowledgments}
The authors thank A.V. Anisovich for useful discussions.

The paper was supported by grant RSF 16-12-10267.

\appendix
\section{Baryons in terms of diquarks}\label{appA}
\subsection{Proton: $N^{+}(uud)$, $I=\frac12$}
Proton $N^{\frac12\frac12,\frac12J_z}$:
\bea\label{19}
p^{\uparrow}&=&\frac{1}{3\sqrt 2}u^{\uparrow}A^{10,10}_{ud}-\frac{1}{3}u^{\downarrow}A^{10,11}_{ud}-\frac{1}{3}d^{\uparrow}A^{11,10}_{uu}+\nn\\
&+&\frac{\sqrt 2}{3}d^{\downarrow}A^{11,11}_{uu}+\frac{1}{\sqrt 2}u^{\uparrow}S^{00,00}_{ud},\\
p^{\downarrow}&=&-\frac{1}{3\sqrt 2}u^{\downarrow}A^{10,10}_{ud}+\frac{1}{3}u^{\uparrow}A^{10,1-1}_{ud}+\frac{1}{3}d^{\downarrow}A^{11,10}_{uu}-\nn\\
&-&\frac{\sqrt 2}{3}d^{\uparrow}A^{11,1-1}_{uu}-\frac{1}{\sqrt 2}u^{\downarrow}S^{00,00}_{ud}.
\eea

Proton $N^{'\frac12\frac12,\frac12J_z}$:
\bea\label{21}
p^{'\uparrow}&=&-\frac{1}{3\sqrt 2}u^{\uparrow}A^{10,10}_{ud}+\frac{1}{3}u^{\downarrow}A^{10,11}_{ud}+\frac{1}{3}d^{\uparrow}A^{11,10}_{uu}-\nn\\
&-&\frac{\sqrt 2}{3}d^{\downarrow}A^{11,11}_{uu}+\frac{1}{\sqrt 2}u^{\uparrow}S^{00,00}_{ud},\\
p^{'\downarrow}&=&\frac{1}{3\sqrt 2}u^{\downarrow}A^{10,10}_{ud}-\frac{1}{3}u^{\uparrow}A^{10,1-1}_{ud}-\frac{1}{3}d^{\downarrow}A^{11,10}_{uu}+\nn\\
&+&\frac{\sqrt 2}{3}d^{\uparrow}A^{11,1-1}_{uu}-\frac{1}{\sqrt 2}u^{\downarrow}S^{00,00}_{ud}.
\eea

Nucleon $N^{\frac12\frac12,\frac32J_z}$:
\bea\label{23}
N^{\frac12\frac12,\frac32\frac32}&=&-\sqrt{\frac13}u^{\uparrow}A_{ud}^{(10,11)}+\sqrt{\frac23}d^{\uparrow}A_{uu}^{(11,11)},\\
N^{\frac12\frac12,\frac32\frac12}&=&-\frac13 u^{\downarrow}A_{ud}^{(10,11)}-\frac{\sqrt{2}}{3}u^{\uparrow}A_{ud}^{(10,10)}+\nn\\
&+&\frac{\sqrt{2}}{3}d^{\downarrow}A_{uu}^{(11,11)}+\frac23d^{\uparrow}A_{uu}^{(11,10)},\\
N^{\frac12\frac12,\frac32-\frac12}&=&-\frac13 u^{\uparrow}A_{ud}^{(10,1-1)}-\frac{\sqrt{2}}{3}u^{\downarrow}A_{ud}^{(10,10)}+\nn\\
&+&\frac{\sqrt{2}}{3}d^{\uparrow}A_{uu}^{(11,1-1)}+\frac23d^{\downarrow}A_{uu}^{(11,10)},\\
N^{\frac12\frac12,\frac32-\frac32}&=&-\sqrt{\frac13}u^{\downarrow}A_{ud}^{(10,1-1)}+\sqrt{\frac23}d^{\downarrow}A_{uu}^{(11,10)}.
\eea

\subsection{Delta: $\Delta^{+}(uud)$, $I=\frac32$}
Delta $\Delta^{+(\frac32\frac12,\frac12 J_z)}$:
\bea\label{25}
\Delta^{+(\frac32\frac12,\frac12\frac12)}&=&-\frac{1}{3}d^{\uparrow}A^{11,10}_{uu}+\frac{\sqrt2}{3}d^{\downarrow}A^{11,11}_{uu}+\nn\\
&+&\frac{2}{3}u^{\downarrow}A^{10,11}_{ud}-\frac{\sqrt2}{3}u^{\uparrow}A^{10,10}_{ud},\\
\Delta^{+(\frac32\frac12,\frac12-\frac12)}&=&\frac{1}{3}d^{\downarrow}A^{11,10}_{uu}-\frac{\sqrt2}{3}d^{\uparrow}A^{11,1-1}_{uu}-\nn\\
&-&\frac{2}{3}u^{\uparrow}A^{10,1-1}_{ud}+\frac{\sqrt2}{3}u^{\downarrow}A^{10,10}_{ud}.
\eea
Delta $\Delta^{+(\frac32\frac12,\frac32 J_z)}$:
\bea\label{26}
\Delta^{+(\frac32\frac12,\frac32\frac32)}&=&\sqrt{\frac{1}{3}}d^{\uparrow}A^{11,11}_{uu}+\sqrt{\frac{2}{3}}u^{\uparrow}A^{10,11}_{ud},\\
\Delta^{+(\frac32\frac12,\frac32\frac12)}&=&\frac{\sqrt{2}}{3}d^{\uparrow}A^{11,10}_{uu}+\frac{1}{3}d^{\downarrow}A^{11,11}_{uu}+\nn\\
&+&\frac{\sqrt{2}}{3}u^{\downarrow}A^{10,11}_{ud}+\frac{2}{3}u^{\uparrow}A^{10,10}_{ud},\\
\Delta^{+(\frac32\frac12,\frac32-\frac12)}&=&\frac{\sqrt{2}}{3}d^{\downarrow}A^{11,10}_{uu}+\frac{1}{3}d^{\uparrow}A^{11,1-1}_{uu}+\nn\\
&+&\frac{\sqrt{2}}{3}u^{\uparrow}A^{10,1-1}_{ud}+\frac{2}{3}u^{\downarrow}A^{10,10}_{ud},\\
\Delta^{+(\frac32\frac12,\frac32-\frac32)}&=&\sqrt{\frac{1}{3}}d^{\downarrow}A^{11,1-1}_{uu}+\sqrt{\frac{2}{3}}u^{\downarrow}A^{10,1-1}_{ud}.
\eea

\subsection{Lambda: $\Lambda_c^{+}(cud)$, $I=0$}
Lambda $\Lambda_c^{+(00,\frac12 J_z)}$:
\bea\label{27}
\Lambda_c^{+(00,\frac12\frac12)}&=&c^{\uparrow}S^{00,00}_{ud},\\
\Lambda_c^{+(00,\frac12 -\frac12)}&=&c^{\downarrow}S^{00,00}_{ud}.
\eea

\subsection{Sigma: $\Sigma_c^{+}(cud)$, $I=1$}
Sigma $\Sigma_c^{+(10,\frac12J_z)}$:
\bea\label{28}
\Sigma_c^{+(10,\frac12\frac12)}&=&\sqrt{\frac23}c^{\downarrow}A^{10,11}_{ud}-\sqrt{\frac13}c^{\uparrow}A^{10,10}_{ud},\\
\Sigma_c^{+(10,\frac12-\frac12)}&=&-\sqrt{\frac23}c^{\uparrow}A^{10,1-1}_{ud}+\sqrt{\frac13}c^{\downarrow}A^{10,10}_{ud}.
\eea
Sigma $\Sigma_c^{+(10,\frac32J_z)}$:
\bea\label{29}
\Sigma_c^{+(10,\frac32\frac32)}&=&c^{\uparrow}A^{10,11}_{ud},\\
\Sigma_c^{+(10,\frac32\frac12)}&=&\sqrt{\frac13}c^{\downarrow}A^{10,11}_{ud}+\sqrt{\frac23}c^{\uparrow}A^{10,10}_{ud},\\
\Sigma_c^{+(10,\frac32-\frac12)}&=&\sqrt{\frac13}c^{\uparrow}A^{10,1-1}_{ud}+\sqrt{\frac23}c^{\downarrow}A^{10,10}_{ud},\\
\Sigma_c^{+(10,\frac32-\frac32)}&=&c^{\downarrow}A^{10,1-1}_{ud}.
\eea

\subsection{Sigma: $\Sigma_c^{++}(cuu)$, $I=1$}
Sigma $\Sigma_c^{++(11,\frac12J_z)}$:
\bea\label{30}
\Sigma_c^{++(11,\frac12\frac12)}&=&\sqrt{\frac23}c^{\downarrow}A^{11,11}_{uu}-\sqrt{\frac13}c^{\uparrow}A^{11,10}_{uu},\\
\Sigma_c^{++(11,\frac12-\frac12)}&=&-\sqrt{\frac23}c^{\uparrow}A^{11,1-1}_{uu}+\sqrt{\frac13}c^{\downarrow}A^{11,10}_{uu}.
\eea
Sigma $\Sigma_c^{++(11,\frac32J_z)}$:
\bea\label{31}
\Sigma_c^{++(11,\frac32\frac32)}&=&c^{\uparrow}A^{11,11}_{uu},\\
\Sigma_c^{++(11,\frac32\frac12)}&=&\sqrt{\frac13}c^{\downarrow}A^{11,11}_{uu}+\sqrt{\frac23}c^{\uparrow}A^{11,10}_{uu},\\
\Sigma_c^{++(11,\frac32-\frac12)}&=&\sqrt{\frac13}c^{\uparrow}A^{11,1-1}_{uu}+\sqrt{\frac23}c^{\downarrow}A^{11,10}_{uu},\\
\Sigma_c^{++(11,\frac32-\frac32)}&=&c^{\downarrow}A^{11,1-1}_{uu}.
\eea

\end{document}